\documentclass[twocolumn,showpacs,fleqn,nobibnotes]{revtex4}

\usepackage{amsmath}
\usepackage{graphicx}
\usepackage{float}
\usepackage{subfigure}
\usepackage{epsfig}
\newcommand{\rr}{\mbox{\boldmath $r$}}

\def\simge{\mathrel{%
   \rlap{\raise 0.511ex \hbox{$>$}}{\lower 0.511ex \hbox{$\sim$}}}}
\def\simle{\mathrel{
   \rlap{\raise 0.511ex \hbox{$<$}}{\lower 0.511ex \hbox{$\sim$}}}}

\begin{document}
\title{Ultrahigh energy neutrinos and non-linear QCD dynamics}
\author{Magno V.T. Machado $^{a,b,c}$ }

\affiliation{$^{a}$ High Energy Physics Phenomenology Group, GFPAE IF-UFRGS \\ CP 15051, CEP 91501-970, Porto Alegre/RS, Brazil\\
$^{b}$ The Abdus Salam International Centre for Theoretical Physics (ICTP) \\ Strada Costiera 11, 34014 Trieste, Italy \\
$^{c}$ CERN TH Division, CH-1211 Gen\`eve 23, Switzerland }

\begin{abstract}
 The ultrahigh energy neutrino-nucleon cross sections are  computed taking into account different phenomenological implementations of the non-linear QCD dynamics. Based on the color dipole framework, the results for the saturation model supplemented by DGLAP evolution as well as for the BFKL formalism in the geometric scaling regime are presented. They are contrasted with recent calculations using NLO DGLAP  and unified BFKL-DGLAP formalisms.  
\end{abstract}

\maketitle

\section{Introduction}

 The inelastic interaction of neutrinos with nucleons is in general described by the QCD-improved parton model. In ultrahigh energy neutrino interactions one probes a kinematical region which is not accessible in the current collider experiments.  It is known so far that these reactions can be sensitive to the domain for the usual Bjorken variable $x\simeq m_{W,Z}^2/E_{\nu} \sim 10^{-8}$ at $E_{\nu}\sim 10^{12}$ GeV and electroweak boson virtualities $Q^2\sim m_{W,Z}^2\approx 10^{3}$ GeV$^2$, where $m_{W,Z}$ are the boson masses. In contrast, in the current electron-proton collider DESY HERA the interactions are probed in the typical kinematical range $x\simge 10^{-5}$ and $0 \leq Q^2\simle 10^3$ GeV$^2$ (the variables $x$ and photon virtuality $Q^2$ are correlated, with small $x$ data associated to low virtualities). There, the measurements of the deep inelastic $ep$ scattering structure functions have put important constraints on parton distributions, mainly sea-quarks and gluons, in the small $x$ region. In perturbative QCD these distributions are expected to grow as $x$ decreases, which has been strongly confirmed by HERA data. Therefore, in order to provide accurate predictions for the ultrahigh energy neutrino-nucleon cross sections, a precise extrapolation of the structure functions to the small $x$ and large $Q^2$ region probed in these interactions is needed.

In the usual QCD-improved parton model, the power growth of the parton distributions functions in very high neutrino energies lead consequently to a power increasing of the neutrino-nucleon total cross section, which in turn implies in a violation of unitarity at high energies. This issue is an outstanding theoretical challenge, which has produced intense work towards a complete understanding of the non-linear (higher twist) QCD corrections in the high parton density regime at fixed impact parameter (for a recent review, see  \cite{armestoreview} and references therein). At present, one has reliable estimates for the kinematical domain where non-linear (saturation) effects should  play an important role. Namely, below a typical transverse  momentum scale $Q_{\mathrm{sat}}^2(x)\sim x^{-\lambda}$, the so-called saturation scale \cite{QSATDEF}, the growth of parton (gluon) density is tamed towards the black-disk limit of the target. This slow down of the gluon distribution function prevents the total cross section violating unitarity. An important prediction of the non-linear QCD approaches is the property of geometric scaling \cite{SGK} below saturation scale, where the gluon density (and cross sections) scales with the saturation momentum, which grows with $1/x$ as a power. It has also been shown that such scaling survives even above the saturation regime \cite{IANCUGEO,STASTOGEO,MUNIERWALLON}: the kinematical region has been estimated to be $Q_{\mathrm{sat}}^2 (x)<Q^2<Q_{\mathrm{sat}}^4 (x)/\Lambda^2_{\mathrm{QCD}}  $ relying on the BFKL equation \cite{BFKL} in the presence of saturation \cite{IANCUGEO}. This feature has important consequences in the underlying QCD dynamics at the nucleus-nucleus accelerators RHIC and LHC and should pose strong constraints to the cross sections beyond current accelerator energies. 
Concerning high energy neutrino cross sections, there are estimates constrained by HERA data based on DGLAP \cite{DGLAP} and effects of BFKL \cite{BFKL} linear evolutions \cite{DKRS,GQRS,RSSSV,GRVNEUT,BASU,KMSNEUT} as well as implementations of non-linear QCD corrections \cite{KUTAK}. The linear NLO DGLAP and unified BFKL-DGLAP results are found to be  consistent with each other presenting somewhat small  deviations \cite{KMSNEUT}. An interesting analytical calculation using an approximate DGLAP solution with initial conditions from a soft non-perturbative model has been recently reported in Ref. \cite{FIORE}. Saturation effects are shown to be quite small in recent DGLAP analysis \cite{BASU}, whereas it was found a reduction in the cross section by a factor 2 at $E_{\nu}\sim 10^{12}$ GeV  considering the saturation model or screening effects via Balitsky-Kovchegov equation \cite{KUTAK}. Recently, it has been claimed in Ref. \cite{JAMAL} that the geometric scaling property can lead to an enhancement of the neutrino-nucleon total cross section by an order of magnitude compared to the leading twist cross section and would lead it further to its unitarization.  


In this work one presents estimates for the ultrahigh energy neutrino-nucleon cross sections, focusing on the saturation models which encode the physics of non-linear QCD evolution in a phenomenological way. The main point is those models provide a suitable dynamical extrapolation to very high energies/virtualities in very contrast with usual DGLAP based approaches, which are not a priori valid in that regime and are dependent on quark and gluon distributions also unknown (no constraint from accelerator data) in that kinematical limit.  Based on the color dipole framework, first one presents the result for the saturation model \cite{GBW} supplemented by QCD evolution \cite{BGBK}. The original saturation model \cite{GBW} was proposed for either low values of $Q^2 \simle 150$ GeV$^2$ , whereas high energy neutrino interactions are sensitive to large virtualities $Q^2\sim m_{W,Z}^2$. Therefore, it is necessary to  perform calculations with complete QCD evolution. Furthermore, one computes  estimates of the ultrahigh neutrino-nucleon cross section through an implementation of the dipole cross section considering BFKL formalism in the geometric scaling region \cite{IIM}.In the last section, the numerical results for both cases are shown and contrasted with recent NLO DGLAP estimates and with unified BFKL-DGLAP formalism without saturation effects.
 
\section{Neutrino-nucleon cross section}

 Deep inelastic neutrino scattering can proceed via  $W^{\pm}$ or $Z^0$ exchanges. The first case corresponds to charged current (CC) and the second one to the neutral current (NC) interactions, respectively.  The standard kinematical variables which describe the processes above are given by, $s=2\,m_N E_{\nu}$, $Q^2=-q^2$, $x=\frac{Q^2}{2p\cdot q}$ and $y=\frac{p \cdot q}{m_N E_{\nu}}$. Here, $m_N$ is the nucleon mass, $E_{\nu}$ labels the neutrino energy and $p$ and $q$ are the four 
momenta of the nucleon and of the exchanged boson, respectively.  The usual Bjorken variable and the momentum transfer are usually labeled as $x$ and  $Q^2$, whereas $s$ is the total centre-of-mass energy squared. One assumes an isoscalar target $N=(p+n)/2$, which is a good approximation for the present purpose.

At high energies a quite successful framework to describe QCD interactions is provided by the color dipole formalism \cite{DIPOLEPIC}, which allow an all twist computation  (in contrast with the usual leading twist approximation) of the structure functions. The physical picture of the interaction is the deep inelastic scattering  at low $x$  viewed as the result of the interaction of a color $q \bar{q}$ dipole which the gauge bosons fluctuate to with the nucleon target. The interaction is modeled via the dipole-target cross section, whereas the boson fluctuation in a color dipole is given by the corresponding wave function. The DIS structure functions for neutrino-nucleon scattering in the dipole picture read as \cite{KUTAK,JAMAL},
\begin{eqnarray}
F_{T,L}^{\mathrm{CC,NC}}(x,Q^2) = \frac{Q^2}{4\,\pi^2}\, \int d^2 \rr \,\int_0^1 dz \,
| \psi^{W^{\pm},Z^0}_{T,L}\,|^2\,\sigma_{dip}\,, 
\label{FSDIP}
\end{eqnarray}
where  $\rr$ denotes the transverse size of the color  dipole, $z$ the 
longitudinal momentum fraction carried by a quark and  $\psi^{W,Z}_{T,L}$ are proportional to the wave functions of the (virtual) charged or neutral gauge bosons corresponding to their  transverse or longitudinal polarization ($F_2=F_T+F_L$). Explicit expressions for the boson ($W^{\pm}$ and $Z^0$) wave functions squared  are as follows \cite{KUTAK},
\begin{eqnarray}
| \psi^{W^{\pm}}_{T}(r,z,Q^2)|^2 & = & \frac{6}{\pi^2}\,[z^2+(1-z)^2]\,\varepsilon^2 \,K_1^2(\varepsilon \rr)\,, \\
| \psi^{W^{\pm}}_{L}(r,z,Q^2)|^2 & = & \frac{24}{\pi^2}\,z^2(1-z)^2 Q^2\,K_0^2(\varepsilon \rr)\,, \\ 
|\psi^{Z^0}_{T}(r,z,Q^2)|^2 & = & \frac{3}{2\,\pi^2}\,K_W\,[z^2+(1-z)^2]\,\varepsilon^2\, K_1^2(\varepsilon \rr),\\
| \psi^{Z^0}_{L}(r,z,Q^2)|^2 & = & \frac{6}{\pi^2}\, K_W\,z^2(1-z)^2 Q^2\,K_0^2(\varepsilon \rr)\,,
\end{eqnarray}  
where one defines the auxiliary variable $\varepsilon= z(1-z)\,Q^2$ and  $K_{0,1}(x)$ are the Mc Donald's functions. Here, $K_{W}=(L_u^2+L_d^2+R_u^2+R_d^2)$ and the the chiral couplings are  expressed as functions of the Weinberg angle $\theta_W$ as follows,
\begin{eqnarray}
L_u & = & 1-{4\over 3} \sin^2\theta_W\,, \hspace{1cm} L_d=-1+{2\over 3} \sin^2\theta_W\,, \\
R_u & = & -{4\over 3} \sin^2\theta_W\,, \hspace{1.4cm} R_d={2\over 3} \sin^2\theta_W\,.
\label{quiralcoup}
\end{eqnarray}

Following Ref. \cite{KUTAK}, here one considers  only four flavors ($u,d,s,c$) and assumes them massless, whereas it has been shown \cite{KMSNEUT} that  heavy quarks $(b,t)$ give relatively small contribution. Moreover, the color dipoles  contributing  to Cabibbo favored transitions are  $ u \bar d \, (d \bar u)$,  
$ c \bar s \,(s \bar c)$ for CC interactions  and $ u \bar u, d \bar d, c \bar c, s \bar s$ for NC interactions.  In our further numerical calculations using the dipole framework, we use the structure function $F_3^{\mathrm{CC,NC}}$ given by the usual LO DGLAP expressions \cite{NUDGLAP}, despite their contributions to be  small for the present purpose. 

The total CC (NC) neutrino-nucleon cross sections as a function of the neutrino energy are given by the integration over available phase space at the given neutrino energy. They read as, 
\begin{eqnarray}
\sigma^{\mathrm{CC,NC}}_{(\nu,\,\bar{\nu})}(E_{\nu})=\int _{Q_{\mathrm{min}}^2}^{s} dQ^2\int_{Q^2/s}^1 dx\, \frac{1}{xs}\,
\frac{\partial^2 \,\sigma_{(\nu,\,\bar{\nu})}^{\mathrm{CC,NC}}}{\partial x\,\partial y}
\label{sigmatot}
\end{eqnarray}
where $y=Q^2/(xs)$ and a minimum value $Q_{\mathrm{min}}^2$ (of order of a few GeV's) on $Q^2$ is introduced in order to stay in the deep inelastic region.  

In the present work we are interested in the ultrahigh energy neutrinos ($E_{\nu}\gg 10^{7}$ GeV), where the valence quark contribution stays constant  and physics is driven  by sea quark contributions. In this region, we will use the available phenomenology on non-linear QCD evolutions to study saturation effects in the neutrino-nucleon cross section. 

The dipole cross section $\sigma_{dip}$, describing the dipole-nucleon
interaction, is substantially affected by saturation effects at dipole sizes $\rr \simge 1/Q_{\mathrm{sat}}$.
 Here, we  follow the  saturation model \cite{GBW}, which
interpolates between the small and large dipole configurations, providing
color transparency behavior, $\sigma_{dip}\sim \rr^2$, as $\rr \rightarrow
0$ and constant behavior, $\sigma_{dip}\sim \sigma_0$, at large
dipoles. The transition is rendered by the saturation phenomenon.   The
parametrisation for the dipole cross section takes the eikonal-like  form,
\begin{eqnarray} \sigma_{dip} (\tilde{x}, \,\rr^2)   =   \sigma_0 \,
\left[\, 1- \exp \left(-\frac{\,Q_s^2(x)\,\rr^2}{4} \right) \, \right]\, (1-\tilde{x})^7 , 
\label{gbwdip}
\end{eqnarray} 
where the saturation scale $Q_s^2(x)   =   \left( \frac{x_0}{\tilde{x}}
\right)^{\lambda}$ defines the onset of the saturation effects. As known, most of the  contribution to ultrahigh energy neutrino cross section comes from very small $x \approx m_{W,Z}^2/2m_N E_{\nu}$ and therefore  a sizeable part of the contributions is in the geometric scaling region or even in the saturation region \cite{JAMAL}. The parameters ($\sigma_0$, $\lambda$ and $x_0$)  were obtained from a fit to the HERA data ~\cite{GBW}. The  variable $\tilde{x}= x\,( \, 1+ 4m_f^2/Q^2)$ gives a suitable  transition to the photoproduction region.  The saturation  model has been used to compute neutrino-nucleon cross section in Ref. \cite{KUTAK}. There, it was shown  that the saturation model does not include complete DGLAP QCD evolution at high virtualities and tends to be therefore lower than other calculations, probably meaning that it is rather incomplete and inaccurate in the high $Q^2$ region.

Recently, a new implementation of the model including QCD evolution
\cite{BGBK} (labeled BGBK model) has appeared. Now, the dipole cross section depends on the gluon distribution as, 
\begin{eqnarray}
 \sigma_{dip}   =  \sigma_0 \, \left[\, 1- \exp
\left(-\frac{\,\pi^2\,\rr^2\,\alpha_s(\mu^2)\,\tilde{x}\,G(\tilde{x},\mu^2)}{3\,\sigma_0}
\right) \, \right]\,,
\label{bgkdip} 
\end{eqnarray} 
where the initial condition at $\mu^2=1$ GeV$^2$ is $x\,G = A_g\,x^{-\lambda_g}\,(1-x)^{5.6}$ and  $\mu^2=C/\rr^2 + \mu_0^2$. The phenomenological parameters are
determined from a fit to small $x$ HERA  data. The function $G(x,\mu^2)$ is evolved with the
leading order DGLAP evolution equation for the gluon density with initial
scale $Q_0^2=1$ GeV$^2$. The improvement preserves the main features of the
low-$Q^2$ and transition regions, while providing QCD evolution in the
large-$Q^2$ domain. 

Despite the saturation model to be very successful in describing HERA data, its functional form is only an approximation motivated by the Glauber-Mueller formula, which it does not include impact parameter dependence. On the other hand, an analytical expression for the dipole cross section can be obtained within the BFKL formalism. Currently, intense theoretical studies has been performed towards an understanding of the BFKL approach in the border of the saturation region \cite{IANCUGEO,MUNIERWALLON}. In particular, the dipole cross section has been calculated in both LO \cite{BFKL} and NLO BFKL \cite{NLOBFKL} approach in the geometric scaling region \cite{BFKLSCAL}. It reads as,
\begin{eqnarray}
\sigma_{dip}= \sigma_0 \,\left[\rr^2 Q_{\mathrm{sat}}^2(x)\right]^{\gamma_{\mathrm{sat}}}\,\exp\left[ -\frac{\ln^2\,\left(\rr^2 Q_{\mathrm{sat}}^2\right)}{2\,\beta \,\bar{\alpha}_sY}\right]\,,
\label{sigmabfkl}
\end{eqnarray}
where $\sigma_0  =2\pi R_p^2$ ($R_p$ is the proton radius) is the overall normalization and the power $\gamma_{\mathrm{sat}}$ is the (BFKL) saddle point in the vicinity of the saturation line $Q^2= Q_{\mathrm{sat}}^2(x)$ (the anomalous dimension is defined as $\gamma = 1- \gamma_{\mathrm{sat}}$). As usual in the BFKL formalism, $\bar{\alpha}_s=N_c\,\alpha_s/\pi$, $\beta \simeq 28\,\zeta (3)$ and the notation $Y=\ln (1/x)$. The quadratic diffusion factor in the exponential gives rise to the scaling violations.  Eq. (\ref{sigmabfkl}) has been used in Ref. \cite{JAMAL} to show an estimation of the enhancement factor (about a factor 10) in ultrahigh neutrino-nucleon cross section in contrast with the usual leading twist calculations.    

The dipole cross section  in Eq. (\ref{sigmabfkl}) does not include an extrapolation from the geometric scaling region to the saturation region. This  has been recently implemented in Ref. \cite{IIM}, where the dipole amplitude  ${\mathcal N} (x,\rr)=\sigma_{dip}/2\pi R_p^2$ was constructed to smoothly interpole between the  limiting behaviors analytically under control: the solution of the BFKL equation
for small dipole sizes, $\rr\ll 1/Q_{\mathrm{sat}}(x)$, and the Levin-Tuchin law \cite{LTLAW}
for larger ones, $\rr\gg 1/Q_{\mathrm{sat}}(x)$. A fit to the structure function $F_2(x,Q^2)$ was performed in the kinematical range of interest, showing that it is  not very sensitive to the details of the interpolation. The dipole cross section was parametrized as follows,
\begin{eqnarray}
\sigma_{dip} = \sigma_0\, \left\{ \begin{array}{ll} 
{\mathcal N}_0\, \left(\frac{\rr\, Q_{\mathrm{sat}}}{2}\right)^{2\left(\gamma_{\mathrm{sat}} + \frac{\ln (2/\rr Q_{\mathrm{sat}})}{\kappa \,\lambda \,Y}\right)}\,, & \!\!(\rr Q_s \le 2)\,,\\
 1 - \exp^{-a\,\ln^2\,(b\,\rr\, Q_{\mathrm{sat}})}\,,  & \!\!(\rr Q_s  > 2)\,, 
\end{array} \right.\nonumber
\label{CGCfit}
\end{eqnarray}
where the expression for $\rr Q_{\mathrm{sat}}(x)  > 2$  (saturation region)   has the correct functional
form, as obtained either by solving the Balitsky-Kovchegov (BK) equation \cite{BK}, 
or from the theory of the Color Glass Condensate (CGC) \cite{CGCREVIEW}. Hereafter, we label the model above by CGC. The coefficients $a$ and $b$ are determined from the continuity conditions of the dipole cross section  at $\rr Q_{\mathrm{sat}}(x)=2$. The coefficients $\gamma_{\mathrm{sat}}= 0.63$ and $\kappa= 9.9$  are fixed from their LO BFKL values. In our further calculations it will be used the parameters $R_p=0.641$ fm, $\lambda=0.253$, $x_0=0.267\times 10^{-4}$ and ${\mathcal N}_0=0.7$, which give the best fit result. A large $x$ threshold factor $(1-x)^5$ will be also considered.

\section{Results and discussion}

\begin{figure}[t] 
\begin{tabular}{cc} 
\epsfig{file=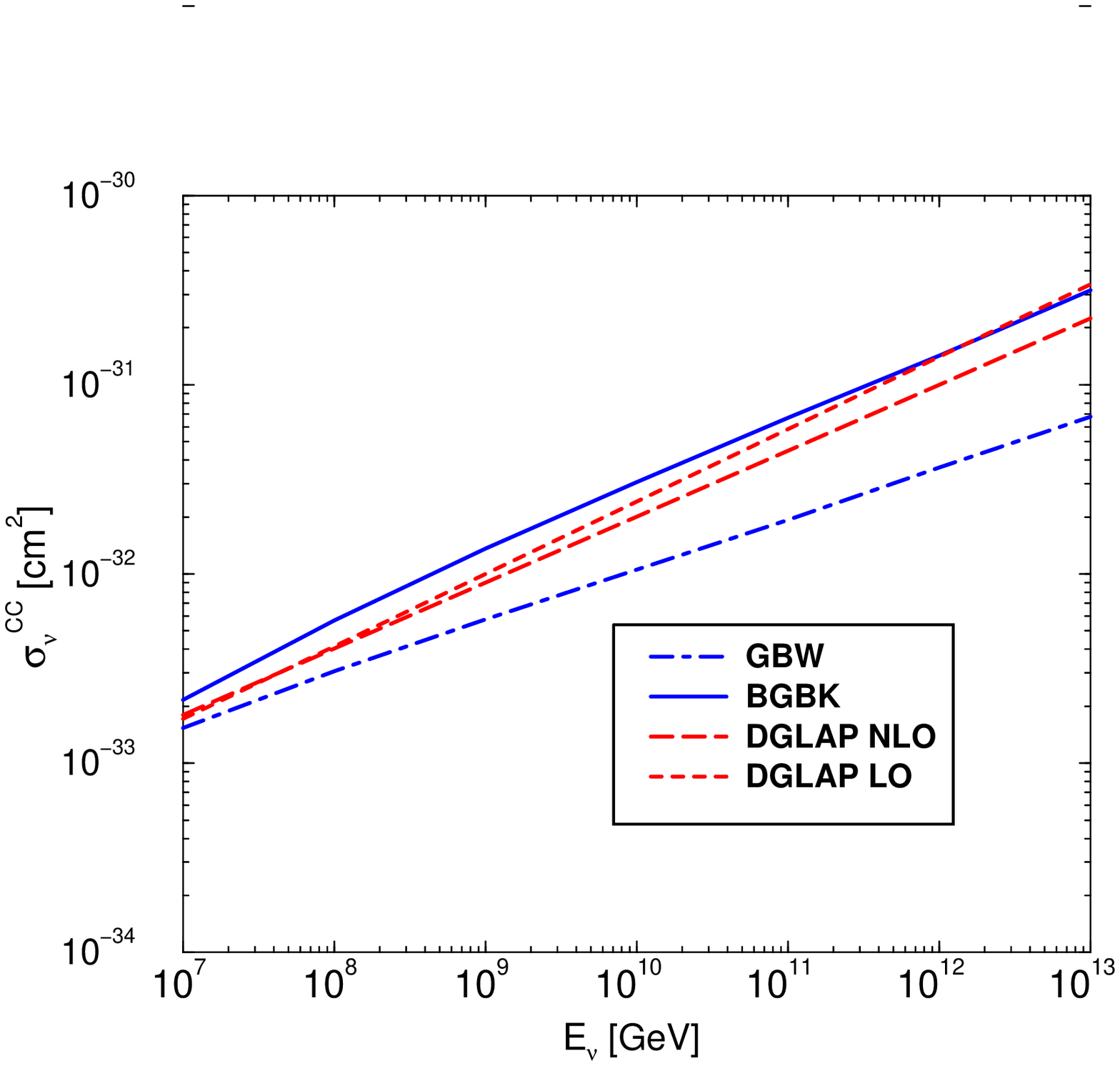,width=43mm,height=55mm} & \epsfig{file=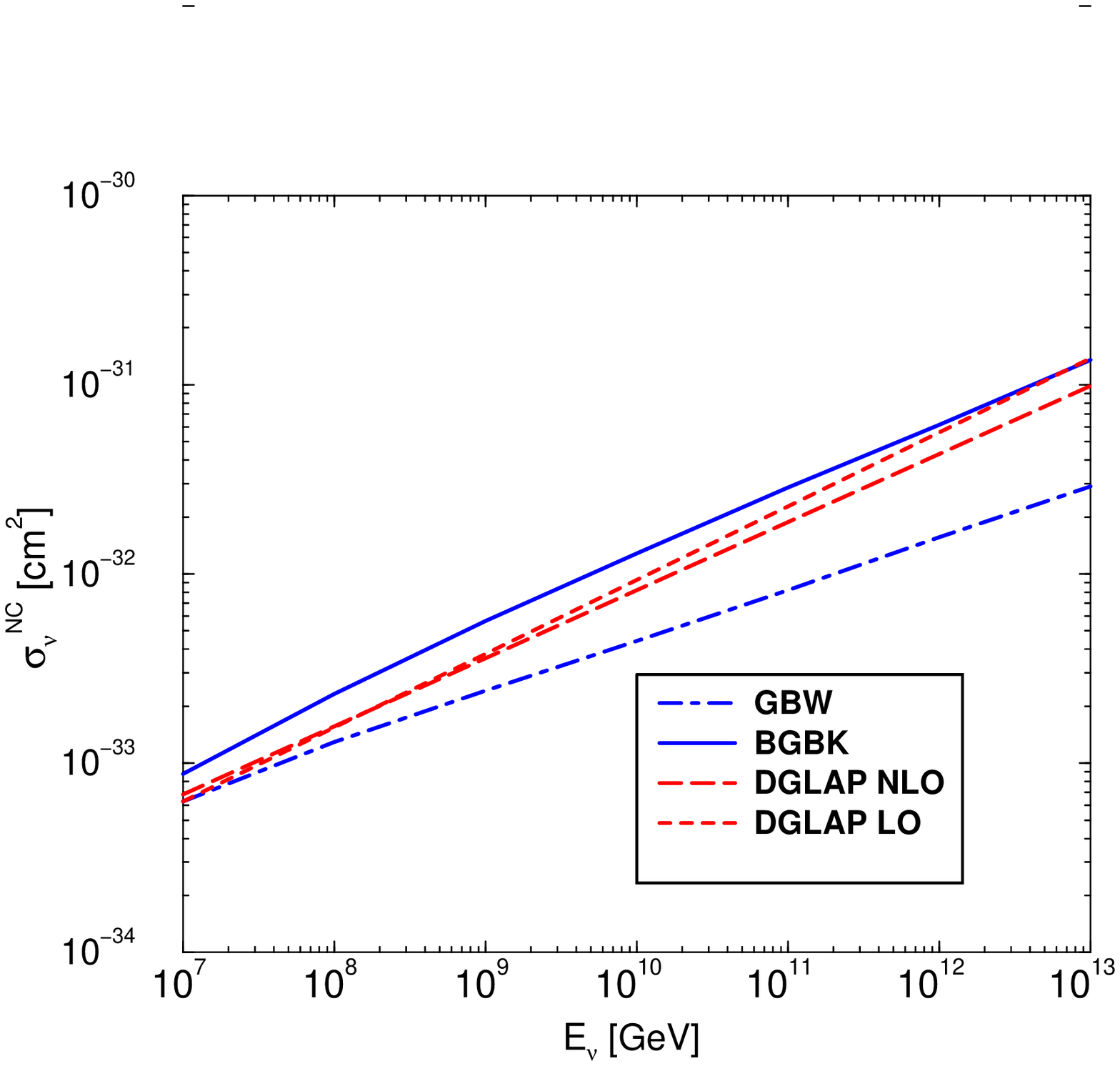,width=43mm,height=55mm} \\ 
	(a) & (b)
\end{tabular} 
\caption{The $\nu \,N$  charged current (a) and  neutral current (b) as a function of neutrino energy. The curves correspond to the results for  the GBW saturation model (dot-dashed curves) and the BGBK model (solid lines). The LO and NLO DGLAP results are represented by dashed and long-dashed curves, respectively.}
\label{fig:1}
 \end{figure}

 In what follows, one  computes the ultrahigh energy CC and NC neutrino-nucleon cross sections using the different models containing saturation effects. In Fig. \ref{fig:1} one presents the results  for the saturation model (GBW), the BGBK model and also   one includes a DGLAP  calculation. The result for the saturation model (dot-dashed lines)  has been recently computed in Ref. \cite{KUTAK}, which we reproduce here. The introduction of the DGLAP evolution in the saturation model enhances the cross section by a factor 3 or even more at $E_{\nu} \sim 10^{12}$ GeV. This is shown by the BGBK curves (solid lines). It should be noticed that the power growth for saturation model is milder than the BGBK model and it gives an upper limit for the role played by saturation effects in the cross section. It should be emphasized this assertion concerns to a comparison between phenomenological saturation model and its  version including DGLAP evolution. Namely, other non-linear QCD approaches can give a larger amount of saturation in the same kinenatical regime and the phenomenological saturation model probably is inaccurate in this region. For sake of completeness, the DGLAP results (LO and NLO) are also presented. For this purpose, we follow the calculations in Ref. \cite{BASU}, where a detailed analysis of the NLO corrections to neutrino-nucleon cross sections was performed. The BGBK model has an energy dependence at ultrahigh energy neutrinos similar to the NLO DGLAP (long-dashed lines) for $E_{\nu} \simge 10^{9}$ GeV.

\begin{figure}[t] 
\begin{tabular}{cc} 
\epsfig{file=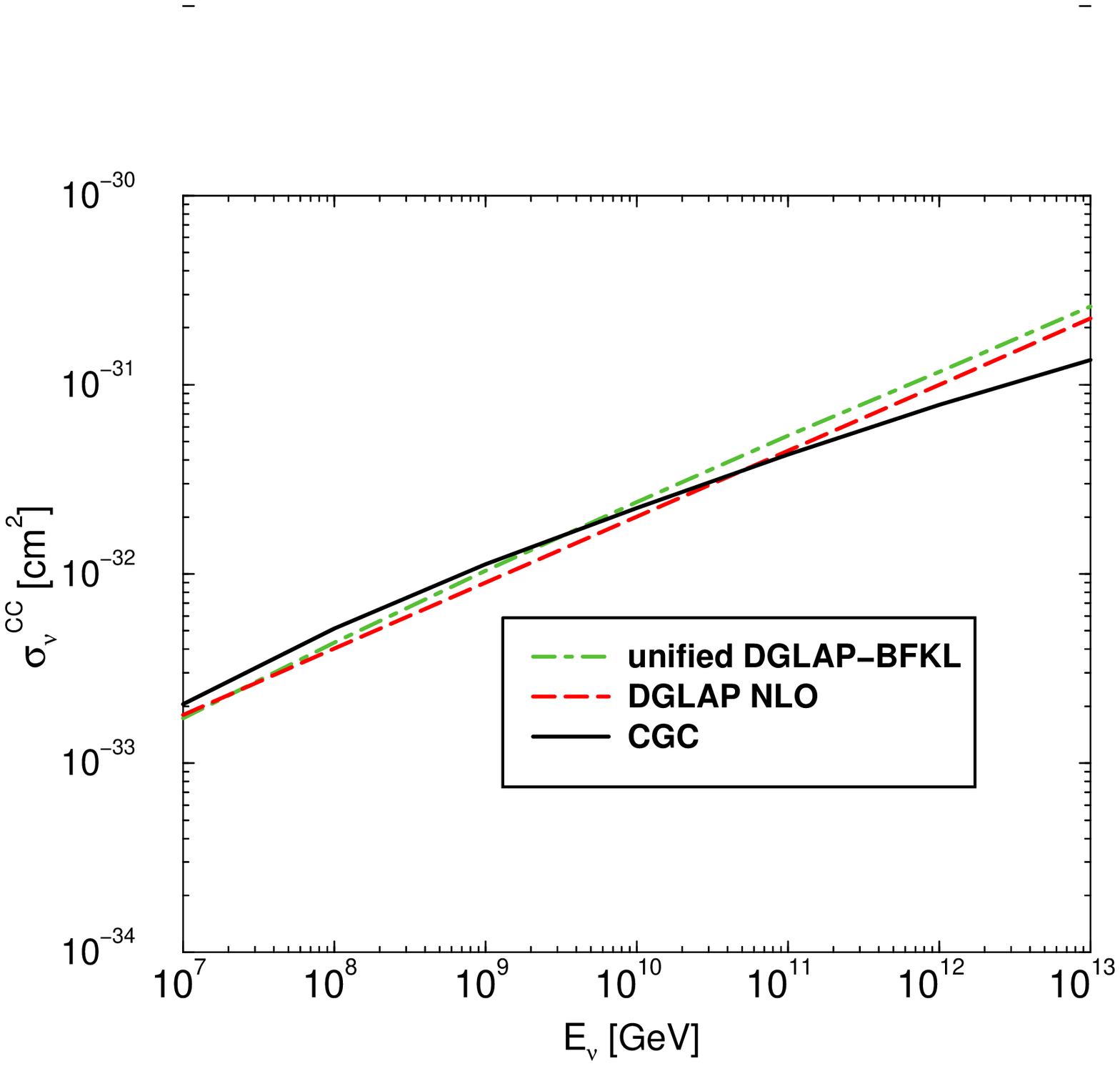,width=43mm,height=55mm} & \epsfig{file=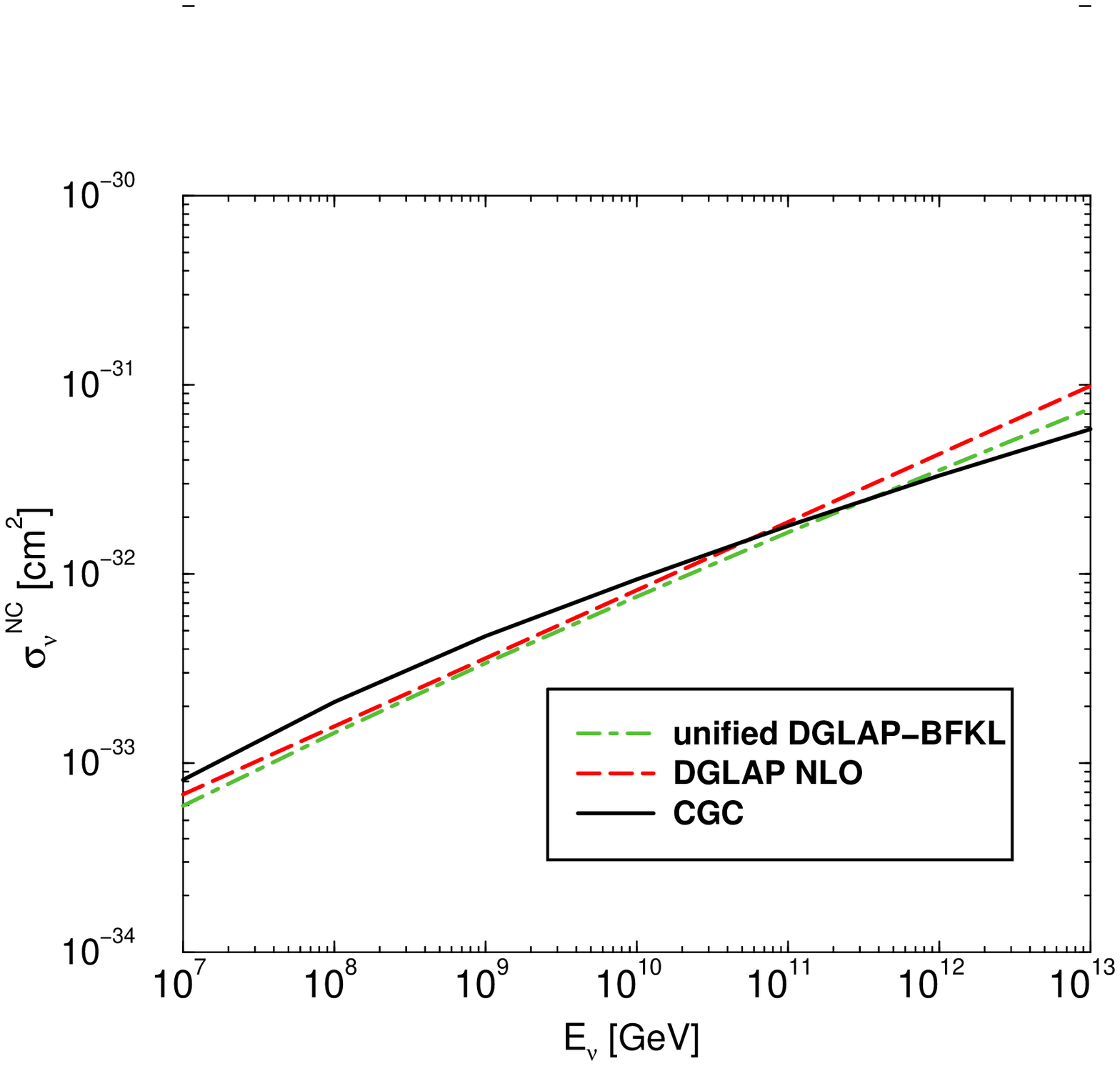,width=43mm,height=55mm} \\ 
(a) & (b) 
\end{tabular} 
\caption{The $\nu \,N$  charged current (a) and  neutral current (b) as a function of neutrino energy. The solid curves correspond to the results for  the CGC model. The NLO DGLAP result is represented by the dashed and the unified BFKL-DGLAP formalism by the dot-dashed curves, respectively.} 
\label{fig:2}
 \end{figure}

In order to investigate the BFKL physics in the geometric scaling region, we use the recent fit to the dipole cross section presented in Ref. \cite{IIM} (the CGC model), which encodes the BFKL formalism in the geometric scaling regime and an interpolation to the saturation domain, as presented in the last section. It should be noticed that an analysis including DGLAP evolution in the fit would be timely. However, as the CGC model produces an ``effective'' anomalous dimension, i.e. $\gamma_{\mathrm{eff}}\,(x,\rr Q_{\mathrm{sat}}) \equiv \gamma_{sat} + (\kappa \lambda Y)^{-1}\ln \,(4/\rr^2Q_{\mathrm{sat}}^2)$ \cite{IIM} far from the geometric scaling region $\rr^2 Q_{\mathrm{sat}}^2(x) \ll 1$, which is closer to the DGLAP one,  the estimate presented here would be not strongly spoiled by QCD evolution. The results for the CGC model (solid lines) is presented in Fig. \ref{fig:2}, contrasted with  the  NLO DGLAP calculation (dashed lines)  and also with the unified BFKL-DGLAP formalism (dot-dashed lines), which embodies non-leading $\ln (1/x)$ contributions \cite{KMSNEUT}.  The latter incorporates both the $\ln(1/x)$ BFKL resummation and the complete LO DGLAP evolution, with the dominant non-leading $\ln(1/x)$ contributions resummed to all orders. The result found for CGC model is quite similar to  both DGLAP and unified BFKL-DGLAP formalisms in the range of energy considered here. It is not observed any enhancement in the neutrino-nucleon cross section in comparison with the leading twist DGLAP calculation, even at very high energies of order $E_{\nu}\sim 10^{12}$ GeV. This fact is in disagreement with Ref. \cite{JAMAL}, where it has been shown that the geometric scaling lead to an enhancement of the neutrino-nucleon total cross section by an order of magnitude versus the  leading twist cross section. We also verify that  at $E_{\nu}\simge  10^{11}$ GeV the power on energy starts to slow down and it should be increasingly smoother (logarithmic growth) at higher energies. This expected change of behavior on energy of the cross section is estimated to be reached at $E_{\nu}\sim 10^{18}$ GeV in Ref. \cite{JAMAL}, suppressing  the cross section compared to the leading twist result.

We can qualitatively understand the result above analyzing the ratio between the dipole cross section for  BFKL  and DGLAP approaches in the geometric scaling regime. As discussed before, the ultrahigh energy neutrino cross section is sensitive to the kinematical region of small $x\sim m_{W,Z}^2/E_{\nu} $ and high $Q^2\sim m_{W,Z}^2$. Concerning DGLAP formalism, in such a region we are in the  double logarithmic approximation (DLA) limit. Concerning the non-linear QCD approaches, at the vicinity of saturation, the DLA saddle point is no longer equal to one, but it was found to be $\gamma_{\mathrm{sat}}^{\mathrm{DLA}}\simeq 1/2$ \cite{IANCUGEO,STASTOGEO,GBMS}. The ratio of the cross section in the geometric scaling region is given by,
\begin{eqnarray}
R_{\mathrm{scal}} \sim \frac{ \left[\rr^2 Q_{\mathrm{sat}}^2 \right]^{ \gamma_{\mathrm{sat}}^{\mathrm{BFKL}}}}{\left[\rr^2 Q_{\mathrm{sat}}^2\right]^{\gamma_{\mathrm{sat}}^{\mathrm{DLA}}}}\simeq \left[\frac{Q_{\mathrm{sat}}^2}{m_{W,Z}^2} \right]^{\left(\gamma_{\mathrm{sat}}^{\mathrm{BFKL}} - \gamma_{\mathrm{sat}}^{\mathrm{DLA}}\right)} \sim {\mathcal O}\,(1)\,,\nonumber
\end{eqnarray}
where we have disregarded the diffusion factor in the BFKL expression and consider $\rr^2\simeq 1/Q^2 \sim 1/m_{W,Z}^2$. The result $R_{\mathrm{scal}}\simeq 1$ leads to similar results between the all twist and leading twist calculation in the geometric scaling region.  In fact, there is room for a slight  suppression since $(\gamma_{\mathrm{sat}}^{\mathrm{BFKL}} - \gamma_{\mathrm{sat}}^{\mathrm{DLA}}) \simeq 0.13$ and $Q_{\mathrm{sat}}^2/m_{W,Z}^2 < 1$. On the other hand, the argument in Ref. \cite{JAMAL}  is exactly true in the comparison between CGC and GBW models. In this case, as we will discuss in what follows, $\gamma_{\mathrm{sat}}^{\mathrm{GBW}}=1$ in the geometric scaling region and a sizeable enhancement between CGC and GBW is really observed.

The qualitative analysis presented above can also help us to understand the enhancement of the BGBK model (with DGLAP evolution) in relation to GBW saturation model. We can think in an effective ``anomalous dimension'' for the saturation model in the geometric scaling regime: from Eq. (\ref{gbwdip}), $\sigma_{dip}^{\mathrm{GBW}} \simeq \rr^2 Q_{\mathrm{sat}}^2/4$ and then one  verifies that $\gamma_{\mathrm{sat}}^{\mathrm{GBW}}=1$. In this case, the ratio now reads as,
\begin{eqnarray}
R_{\mathrm{scal}} = \frac{\sigma_{dip}\,(\mathrm{BGBK})}{\sigma_{dip}\,(\mathrm{GBW})} \sim \left[\frac{Q_{\mathrm{sat}}^2}{m_{W,Z}^2} \right]^{\left(\gamma_{\mathrm{sat}}^{\mathrm{DLA}} - \gamma_{\mathrm{sat}}^{\mathrm{GBW}}\right)} \gg 1\,,
\end{eqnarray}
whereas at lower energies (beyond the geometric scaling regime) the DGLAP anomalous dimension takes the usual value and the ratio is closer to one, as expected. 

As a summary, we have presented the results for the ultrahigh energy neutrino cross section considering the current phenomenology on non-linear QCD dynamics, which play an important role in the correct extrapolation to the regime of very high gluon density. The saturation model with QCD evolution was studied as well as the physics of BFKL dynamics at geometric scaling region. The latter presents suppression in the cross section at $E_{\nu}\simge 10^{12}$ GeV, whereas the BGBK model is consistent with NLO DGLAP calculations. An outstanding advantage in the color dipole framework is the possibility to provide analytical expressions for the extrapolation to ultrahigh energies, towards the black limit of the nucleon target. This cannot be taken into account in a leading twist approach as the QCD-improved  parton model. Then these saturation (all twist) approaches  play an important role, since a precise knowledge about the neutrino  interactions and production rates is essential for estimating background, expected fluxes and detection probabilities \cite{KMSNEUT,STASTO}. In particular, reliable estimates for UHE neutrino cross section are strongly needed for the planned experiments \cite{KUSENKO} to detect them via nearly horizontal air showers in the Earth atmosphere, as discussed in \cite{FIORE}.

\begin{acknowledgments}
 The author is grateful for the warm hospitality and financial support of  the High Energy Group in The Abdus Salam International Centre for Theoretical Physics (ICTP) at Trieste and CERN Theory Division at Geneve, where part of this work was performed. The author also would like to thank Alexei Smirnov (ICTP, Trieste) and Krysztof Kutak (II Institute for Theoretical Physics, Hamburg) for helpful discussions.  This work was partially supported by CNPq and FAPERGS, Brazil.

\end{acknowledgments}


\begin{thebibliography}{99}
\bibitem{armestoreview} N. Armesto, Acta Phys.\ Polon.\ B {\bf 35}, 213 (2004)..

\bibitem{QSATDEF} L.V. Gribov, E.M. Levin and M.G. Ryskin,  {\it Phys. Rep.} 
              {\bf 100}, (1983) 1; A. H. Mueller, {\it Nucl. Phys.} {\bf B558} (1999) 285.

\bibitem{SGK} A.M.  Sta\'sto, K. Golec-Biernat and J. Kwieci\'nski,  {\it Phys. Rev. Lett.} {\bf 86} (2001) 596.   


\bibitem{IANCUGEO} E.~Iancu, K.~Itakura and  L.~McLerran, {\it Nucl. Phys} {\bf A708} (2002) 327.   


\bibitem{STASTOGEO} J.~Kwieci\'nski and A.~M.~Sta\'sto, {\it Phys.\ Rev.\ } {\bf D66} (2002) 014013.

\bibitem{MUNIERWALLON}
S. Munier, S. Wallon, {\it Eur. Phys. J.}  {\bf C30} (2003) 359.


\bibitem{BFKL}
L. N. Lipatov, Sov. J. Nucl. Phys. {\bf 23}, 338 (1976); E. A.
Kuraev, L. N. Lipatov, V. S. Fadin, JETP {\bf 45}, 1999 (1977); I.
I. Balitskii, L. N. Lipatov, Sov. J. Nucl. Phys. {\bf 28}, 822
(1978).

\bibitem{DGLAP} V.N. Gribov and L.N. Lipatov, Sov. J. Nucl. Phys. {\bf 15}, 438 (1972);
G. Altarelli and G. Parisi, Nucl. Phys.  {\bf B126}, 298 (1977);
Yu.L. Dokshitzer, Sov. Phys. JETP {\bf 46}, 641 (1977).

\bibitem{DKRS}
D. A. Dicus, S. Kretzer, W. W. Repko and C. Schmidt, {\it Phys. Lett.} {\bf
B514} (2001) 103.

\bibitem{GQRS}
R. Gandhi, C. Quigg, M.H. Reno and I. Sarcevic, {\it Astropart. Phys.} {\bf 5} (1996) 81; {\it Phys. Rev.} {\bf D58}
(1998) 093009.

\bibitem{RSSSV}
M. H. Reno, I. Sarcevic, G. Sterman, M. Stratmann and W. Vogelsang, hep-ph/0110235.

\bibitem{GRVNEUT} M. Gl\"uck, S. Kretzer and E. Reya, {\it Astropart. Phys.} {\bf 11} (1999) 327. 

\bibitem{BASU} R. Basu, D. Choudhury, S. Majhi, {\it JHEP} {\bf 0210} (2002) 012. 

\bibitem{KMSNEUT}J. Kwieci\'nski, A.D. Martin, A.M. Sta\'sto, {\it Phys. Rev.} {\bf D59} 
(1999) 093002. 

\bibitem{KUTAK} K. Kutak, J. Kwieci\'nski, {\it Eur. Phys. J} {\bf C29} (2003) 521. 

\bibitem{FIORE} R. Fiore {\it et al.}, {\it Phys. Rev.} {\bf D68} (2003) 093010. 

\bibitem{JAMAL} J. Jalilian-Marian, {\it Phys. Rev.} {\bf D68} (2003) 054005.


\bibitem{GBW} K. Golec-Biernat and  M. W\"usthoff,  {\it Phys. Rev.} {\bf D59} (1999) 014017,  {\it ibid.} {\bf D60} (1999) 114023; 

\bibitem{BGBK}
J. Bartels, K. Golec-Biernat, H. Kowalski, {\it  Phys. Rev.} {\bf D66} (2002) 014001.

\bibitem{IIM} E.~Iancu, K. Itakura and S. Munier, Phys.\ Lett.\ B {\bf 590}, 199 (2004).

\bibitem{NUDGLAP}
B. Humpert and W. L. van Neerven,  Nucl. Phys. {\bf B184}, (1981) 225.

\bibitem{DIPOLEPIC}
A. H. Mueller, {\it Nucl. Phys.} {\bf B335} (1990) 115;
N.N. Nikolaev and B.G. Zakharov, {\it Z. Phys.} {\bf C49}
(1991) 607. 

\bibitem{NLOBFKL} V.S.~Fadin and L.N.~Lipatov, 
{\it Phys. Lett.} {\bf B429} (1998) 127;
G. Camici and M. Ciafaloni, {\it Phys. Lett.} {\bf B430}  (1998) 349.

\bibitem{BFKLSCAL} A.H. Mueller and D.N. Triantafyllopoulos, {\it Nucl. Phys.} {\bf B640} (2002) 331; D.N. Triantafyllopoulos, {\it Nucl. Phys.} {\bf B648} (2003) 293; A.H. Mueller, {\it Nucl. Phys.} {\bf A724}  (2003) 223.

\bibitem{LTLAW} E.~Levin and K.~Tuchin, {\it  Nucl. Phys.} {\bf B573} (2000) 833.

\bibitem{BK} I. Balitsky, {\it Nucl. Phys.} {\bf B463} (1996) 99;  Yu.V. Kovchegov, {\it Phys. Rev.} {\bf D60} (1999) 034008; {\it ibid.} {\bf D61} 
(2000) 074018. 

\bibitem{CGCREVIEW} E.~Iancu, R.~Venugopalan, hep-ph/0303204. 

\bibitem{GBMS} K. Golec-Biernat, L. Motyka, A.M. Sta\'sto, {\it Phys. Rev.} {\bf D65} (2002) 074037.

\bibitem{STASTO} A.M. Sta\'sto, Int.\ J.\ Mod.\ Phys.\ A {\bf 19}, 317 (2004).

\bibitem{KUSENKO} A. Kusenko, T.J. Weiler, {\it Phys. Rev. Lett.} {\bf 88} (2002) 161101; A. Kusenko, hep-ph/0203002. 


\end{thebibliography}
\end{document}